\documentclass[epj]{svjour}
\usepackage{graphics}
\usepackage{amssymb}

\usepackage{latexsym}
\usepackage{amsmath}
\usepackage{graphicx}
\usepackage{float}
\usepackage[nooneline]{subfigure}
\begin{document}
\title{Role of the attractive intersite interaction in the extended Hubbard model}

\author{F. Mancini\inst{1} \and F. P. Mancini\inst{2} \and A. Naddeo\inst{3}}
\institute{Dipartimento di Fisica {\it ``E. R. Caianiello''} -
Unit\`a CNISM di Salerno \\Universit\`a degli Studi di Salerno,
Via S.
Allende, I-84081 Baronissi (SA), Italy \and Dipartimento di Fisica and Sezione I.N.F.N.\\
Universit\`a degli Studi di Perugia, Via A. Pascoli, I-06123
Perugia, Italy \and Unit\`a CNISM di Salerno - Dipartimento di
Fisica {\it ``E. R. Caianiello''} \\Universit\`a degli Studi di
Salerno, Via S. Allende, I-84081 Baronissi (SA), Italy}
%
%
\abstract{We consider the extended Hubbard model in the atomic
limit on a Bethe lattice with coordination number $z$. By using
the equations of motion formalism, the model is exactly solved for
both attractive and repulsive intersite potential $V$. By focusing
on the case of negative $V$, i.e., attractive intersite
interaction, we study the phase diagram at finite temperature and
find, for various values of the filling and of the on-site
coupling $U$, a phase transition towards a state with phase
separation. We determine the critical temperature as a function of
the relevant parameters, $U/|V|$, $n$ and $z$ and we find a
reentrant behavior in the plane ($U/|V|$,$T$). Finally, several
thermodynamic properties are investigated near criticality.
} 
\maketitle

\section{Introduction}

Understanding the effects of competing interactions and the
corresponding quantum phase transitions in models of strongly
correlated electron systems is a crucial problem in condensed
matter physics. A thorough comprehension could shed new light on
the behavior of new classes of novel materials, ranging from
inorganic chain compounds and conducting polymers to organic and
high temperature superconductors. In all such systems, long-ranged
Coulomb interactions play an essential role, as it has been
pointed out in Ref. \cite{imada}, where it was shown that the
simple Hubbard model does not offer a model for the
superconductivity in the right order of amplitude. Similarly,
Fr\"{o}lich and Coulomb long-ranged interactions have been
recognized to play an essential role in a realistic multi-polaron
model of high-$T_c$ superconductivity \cite{alexandrov3}.

Here, we study the so-called extended Hubbard model (EHM) \cite
{exhub1}, where a nearest-neighbor interaction $V$ is added to the
original Hubbard Hamiltonian, containing only an on-site
interaction $U$. The intersite interaction can somehow mime
longer-ranged Coulomb interactions. In various applications of the
model, the parameters $U$ and $V$ can represent the effective
interaction couplings taking into account also other interactions
(for instance with phonons). Therefore we assume that $U$ and $V$
could take positive as well as negative values. The EHM has been
deeply investigated by means of various analytical techniques. In
particular, g-ology investigations \cite{exhub3} as well as
renormalization group \cite{exhub4} and bosonization \cite{exhub5}
analysis have been successfully carried out in the weak coupling
regime, while exact diagonalization calculations \cite{exhub6} and
quantum Monte Carlo simulations \cite{exhub7} have been used to
study the intermediate and strong coupling regimes \cite{exhub8}.
The cases of half filling ($n=1$) \cite{exhub7} and quarter
filling ($n=0.5$) \cite{exhub9} have been deeply investigated in
order to establish the boundary line between spin-density wave and
charge-density wave phases, as well as existence of an
intermediate bond-order wave phase. More recently several
systematic studies of the phase diagram of the one-dimensional
extended Hubbard model as a function of all the three parameters
$n$, $U$, $V$ have been performed as well \cite{exhub10},
revealing a very rich structure due to the presence of multiple
first and second order phase transitions, critical points and
interesting reentrant behaviors \cite{exhub11}. The relation with
the experimental findings relative to some manganese compounds
\cite{exhub12} has been investigated in detail together with the
possibility of improving the description of cuprates with respect
to the simple Hubbard model \cite{exhub11}. However, despite of
the above results, such a general case still lacks a detailed
study.

\begin{figure}[t]
\centering\includegraphics[scale=0.3]{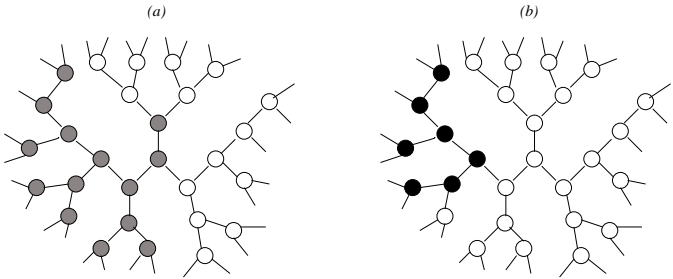}
\caption{\label{fig1} Distribution of the particles along the
Bethe lattice at $n=0.5$ and $T=0$: ($a$) $V<0$ and $U \gtrapprox
2V$; ($b$) $V<0$ and $U \lessapprox 2V$. White, grey and black
squares denote empty, arbitrary spin singly occupied and double
occupied sites, respectively.}
\end{figure}

In order to overcome the difficulties shown by the EHM and to
investigate charge ordering in interacting electron systems, one
may consider a minimal model, allowing one to capture all the
relevant phenomenology while being exactly solved. That is, one
can consider the very narrow band limit, where $t_{ij}\ll U,V$;
for simplicity the classical limit $t_{ij}=0$ can be taken. In
this way one is led to the extended Hubbard model  in the atomic
limit (AL-EHM), which is one of the simplest models apt to
describe charge ordering in interacting electron systems. The
charge ordering has attracted much attention after the discovery
of charge-density waves in high temperature superconductors and
since then the problem of competition and coexistence of this kind
of ordering and other orderings has been widely investigated
\cite{ch1}. Charge ordering is in fact accompanied by
metal-insulator transitions through the growth of charge-order
parameter and the opening of gaps at the Fermi surface.
Furthermore, phenomena such as colossal magnetoresistance in
perovskite-type manganese oxides are observed immediately when the
charge order melts \cite{ch2,ch3}. Recently charge orderings and
charge-density waves have been widely observed in transition metal
compounds \cite{ch4} and also in organic conductors \cite{ch5}. It
is worth noticing that the observed changes in transport, optical,
dielectric and magnetic properties at or near charge-order
transitions have spurred a lot of interest on the problem of their
control through the charge-order transitions. Thus, in this
context, a deeper understanding of the nature of charge-order
transitions could be relevant as well as of the whole structure of
the phase diagrams. Another aspect which is very interesting and
deserves further investigations is the phase separation phenomenon
\cite{phasesep1,pawlowski1,phasesep1a}; it has been investigated
in some detail in the AL-EHM context by means of Monte Carlo
simulations and Hartree-Fock approximation in Ref.
\cite{phasesep2} and the results have been compared with the
experimental ones obtained for the organic conductor
$(DI-DCNQI)_{2}Ag$ \cite{phasesep3}. The phase separation
phenomenon is characterized by a ground state macroscopically
inhomogeneous  with different spatial regions with different
average charge densities. Typically, it occurs for attractive
on-site and intersite interactions, so that the electrons cluster
together in order to gain the maximum negative potential energy.
In the extended Hubbard model, two types of phase separated
configurations are possible: a cluster of singly occupied sites or
a cluster of doubly occupied sites, as shown in Fig. \ref{fig1}.
A cluster of doubly occupied sites is favored when the on-site and
intersite interactions are both attractive, whereas cartoon ($b$)
would correspond to large positive $U$ and negative $V$.

The critical behavior of the AL-EHM has been studied by many
authors, mainly in the one dimensional case
\cite{sf1,sf2,sf3,sf4,sf5,sf6,sf7,sf8,sf9,2}, and a staggered
charge order has been found irrespectively of the method of
analysis. Very recently, a comprehensive and systematic analysis
of the one dimensional AL-EHM has been carried out both at $T=0$
and finite temperature in the whole parameters space and various
types of long-range charge ordered states have been observed
\cite{mancini2}. A study of the two-dimensional version of the
model has been performed \cite{pawlowski1,pawlowski}, and very
recently the same model on a Bethe lattice with general
coordination number $z$ has been investigated \cite{4}.

The study of the extended Hubbard model with attractive as well as
repulsive on-site and intersite interactions could also shed new
light on the properties of systems of ultracold fermionic atoms
loaded in optical lattices, such as fermionic superfluid
properties in a two dimensional lattice \cite{volcko}, and the
possibility of the supersolid state \cite{jaksch}.

In this paper we study the AL-EHM on a Bethe lattice with
coordination number $z$ within the equations of motion framework
\cite{2,1,3}. This formalism allows us to exactly solve the model.
By focusing on the case of attractive $V$, we study the phase
diagram at finite temperature $T$ for different values of the
filling and the coupling $U/\left| V\right|$. By varying the
temperature, we find a region of negative compressibility, hinting
at a transition from a thermodynamically stable to an unstable
phase characterized by phase separation. The critical temperature,
function of the external parameters $U/\left| V\right|$, $n$ and
$z$, shows a reentrant behavior in the plane $\left( U/\left|
V\right| ,T\right)$ at half filling ($n=1$). Although
thermodynamically unstable, the study of the $T<T_c$ phase allows
us to determine the critical value of the on-site interaction
$U_c$ separating the formation of clusters of singly and doubly
occupied sites.

The plan of the paper is as follows: in Sec. \ref{sec_II}, we
briefly review the equations of motion method for the extended
Hubbard model in the atomic limit on a Bethe lattice with
coordination number $z$ and compute the Green's and correlation
functions. We find that they crucially depend only on two
parameters, which can be self-consistently computed allowing us to
determine all the local properties of the system in the case of
general filling $n$. In Sec. \ref{sec_III} we derive the phase
diagram in the planes ($n,T$) and ($U,T$) for a Bethe lattice with
coordination number $z=3$, and we show the behavior of the
relevant thermodynamic quantities. Finally, Sec. \ref{sec_IV} is
devoted to our concluding remarks.

\section{The general model}
\label{sec_II}

The extended Hubbard Hamiltonian in the narrow-band limit on a Bethe lattice
with coordination number $z$ and nearest-neighbor interactions is given by:
\begin{equation}
H=-\mu n(0)+UD(0)+\sum\limits_{p=1}^{z}H^{(p)}.  \label{eq2}
\end{equation}
Here $H^{(p)}$ represents the Hamiltonian of the $p$-th sub-tree
rooted at the central site (0):
\begin{equation}
H^{(p)}=-\mu n(p)+UD(p)+Vn(0)n(p)+\sum\limits_{m=1}^{z-1}H^{(p,m)},
\label{eq3}
\end{equation}
where $p$ ($p=1,\ldots,z$) are the nearest neighbors of the site
(0), also termed the first shell. In turn, $H^{(p,m)}$ describes
the $m$-th sub-tree rooted at the site $(p)$, and so on to
infinity. $U$ and $V$ are the strengths of the local and intersite
interactions, $\mu$ is the chemical potential, $n(i)=n_{\uparrow
}(i)+n_{\downarrow }(i)$ and $D(i)=n_{\uparrow }(i)n_{\downarrow
}(i)=n\left( i\right) \left[ n\left( i\right) -1\right] /2$ are
the charge density and double occupancy operators at a general
site $i$ . As usual, $n_{\sigma }(i)=c_{\sigma }^{\dag
}(i)c_{\sigma }(i)$ with $ \sigma =\left\{ {\uparrow ,\downarrow
}\right\} $ and $c_{\sigma }(i)$ ($ c_{\sigma }^{\dag }(i))$ is
the fermionic annihilation (creation) operator of an electron of
spin $\sigma $ at site $i$, satisfying canonical anticommutation
relations. We adopt the Heisenberg picture $i=\left( \mathbf{
i},t\right)$, and we use the spinorial notation for the fermionic
fields.

The exact solution of the model can be obtained by using the
equations of motion approach in the context of the composite
operator method \cite{5}.  Following this line, the main point is
the choice of the field operators. A convenient operatorial basis
is given by the Hubbard operators, $\xi (i)=[n(i)-1]c(i)$ and
$\eta (i)=n(i)c(i)$, which satisfy the equations of motion:
\begin{equation}
\begin{split}
i\frac{\partial }{\partial t}\xi (i)&=-\mu \xi (i)+zV\xi (i)n^{\alpha }(i) ,\\
i\frac{\partial }{\partial t}\eta (i)&=(U-\mu )\eta (i)+zV\eta
(i)n^{\alpha }(i).
\end{split}
\label{eq4a}
\end{equation}
Hereafter, for a generic operator $\Phi \left( i\right) $ we shall
use the notation $\Phi ^{\alpha
}(i)=\frac{1}{z}\sum\nolimits_{p=1}^{z}\Phi (i,p)$, $ (i,p)$ being
the first nearest neighbors of the site $i$. The Heisenberg
equations \eqref{eq4a} contain the higher-order nonlocal operators
$\xi (i)n^{\alpha }(i)$ and $\eta (i)n^{\alpha }(i)$. By taking
time derivatives of the latter, higher-order operators are
generated. This process may be continued and usually an infinite
hierarchy of field operators is created. However, by noting that
the number $n(i)$ and the double occupancy $D(i)$ operators
satisfy the following algebra
\begin{equation}
\begin{split}
n^{p}\left( i\right) &=n\left( i\right) +a_{p}D(i), \\
D^{p}\left( i\right) &=D(i), \\
n^{p}\left( i\right) D(i)&=2D(i)+a_{p}D(i),
\end{split}
 \label{eq9}
\end{equation}
where $p  \geq 1$ and $a_{p}=2^{p}-2$, it is easy to establish the
following recursion rule:
\begin{equation}
\lbrack n^{\alpha
}(i)]^{k}=\sum\limits_{m=1}^{2z}A_{m}^{(k)}[n^{\alpha }(i)]^{m}.
\label{eq5}
\end{equation}
The coefficients $A_{m}^{(k)}$ are rational numbers, satisfying
the relations $\sum_{m=1}^{2z}A_{m}^{(k)}=1$ and
$A_{m}^{(k)}=\delta _{m,k}$ $(k=1,\cdots ,2z)$ \cite{2}. The
recursion relation (\ref{eq5}) allows one to close the hierarchy
of equations of motion. Thus, one may define a new composite field
operator \cite{4}:
\begin{equation}
\psi (i)=\left( {{\begin{array}{*{20}c} {\psi ^{(\xi )}(i)} \\ {\psi ^{(\eta
)}(i)} \\ \end{array}}}\right) ,
 \label{5a}
\end{equation}
where
\begin{equation}
\psi ^{(\xi )}(i)=
\left( {{\begin{array}{*{20}c} {\xi (i)} \hfill \\ {\xi
(i)[n^\alpha (i)]} \\ \vdots \\ {\xi (i)[n^\alpha (i)]^{2z}} \\ \end{array}}}%
\right) ,\quad \psi ^{(\eta )}(i)=\left( {{\begin{array}{*{20}c} {\eta (i)}
\\ {\eta (i)[n^\alpha (i)]} \\ \vdots \\ {\eta (i)[n^\alpha (i)]^{2z}} \\
\end{array}}}\right)
  \label{eq6}
\end{equation}
satisfy the equations of motion
\begin{equation}
\begin{split}
i\frac{\partial }{\partial t}\psi ^{(\xi )}(i)=[\psi ^{(\xi
)}(i),H]&=\varepsilon ^{(\xi )}\psi ^{(\xi )}(i) \\
i\frac{\partial }{\partial t}\psi ^{(\eta )}(i)=[\psi ^{(\eta
)}(i),H]&=\varepsilon ^{(\eta )}\psi ^{(\eta )}(i).
\end{split}
 \label{eq7}
\end{equation}
Here $\varepsilon ^{(\xi )}$ and $\varepsilon ^{(\eta )}$ are the energy
matrices of rank $(2z+1)\times (2z+1)$ \cite{4} whose eigenvalues are:
\begin{equation}
\begin{split}
E_{m}^{(\xi )}&=-\mu +(m-1)V ,\\
E_{m}^{(\eta )}&=-\mu +U+(m-1)V,
\end{split}
\label{eq10}
\end{equation}
where $m=1,\ldots,(2z+1)$. The model has now been formally solved
since one has found a closed set of eigenoperators and eigenvalues
allowing one to ascertain exact expressions of the retarded
Green's function (GF)
\begin{equation}
 \label{BEHM_8}
 G^{(s)}(t-t')=\theta (t-t')\langle\{\psi
^{(s)}(0,t), {\psi^{s}}^\dag
 (0,t')\}\rangle,
\end{equation}
and, consequently, of the correlation function (CF)
\begin{equation}
\label{BEHM_9}
C^{(s)}(t-t')=\langle \psi^{(s)}(0,t) {\psi^{(s)}}
^\dag (0,t')\rangle.
\end{equation}
Indeed, by means of the equations of motion (\ref{eq7}) one finds
\cite{4}:
\begin{equation}
G^{(s)}(\omega )=\sum_{m=1}^{2z+1}\frac{\sigma ^{(s,m)}}{\omega
-E_{m}^{(s)}+i\delta },  \label{eq13}
\end{equation}
and
\begin{equation}
C^{(s)}(\omega )=\pi \,\sum\limits_{m=1}^{2z+1}\sigma
^{(s,n)}T_{m}^{(s)}\,\delta \left( {\omega -E_{m}^{(s)}}\right) ,
\label{eq20}
\end{equation}
where $s=\xi ,\eta $, $T_{m}^{(s)}=1+\tanh ({\beta
E_{m}^{(s)}/2})$, $\beta = 1/k_{B}T$ and $E_{m}^{(s)}$ are given
in Eq. (\ref{eq10}). The spectral density matrices $\sigma
_{ab}^{(s,n)}$ can be computed by means of the formula:
\begin{equation}
\sigma _{ab}^{(s,n)}=\Omega
_{an}^{(s)}\sum\limits_{c=1}^{2z+1}\left[ { \Omega
_{nc}^{(s)}}\right] ^{-1}I_{cb}^{(s)}.  \label{eq14}
\end{equation}
In Eq. (\ref{eq14}), $\Omega ^{(s)}$ is the $(2z+1)\times (2z+1)$ matrix
whose columns are the eigenvectors of the energy matrix $\varepsilon ^{(s)}$
and $I^{(s)}$ is the $(2z+1)\times (2z+1)$ normalization matrix whose
elements can be written as \cite{4}: $I_{n,m}^{(\xi )}=\kappa
^{(n+m-2)}-\lambda ^{(n+m-2)}$, $I_{n,m}^{(\eta )}=\lambda ^{(n+m-2)}$. Here
the charge correlators $\kappa ^{(k)}$ and $\lambda ^{(k)}$ are defined as
\begin{equation}
\kappa ^{(k)}=\langle {\lbrack n^{\alpha }(0)]^{k}}\rangle ,\quad \lambda
^{(k)}=\frac{1}{2}\langle {n(0)[n^{\alpha }(0)]^{k}}\rangle .  \label{eq17}
\end{equation}
By exploiting the recursion relation (\ref{eq5}), it is not
difficult to show that also $\kappa ^{(k)}$ and $\lambda ^{(k)}$
obey to similar recursion relations, limiting their computation to
the first $2z$ correlators \cite{4}. However, the knowledge of the
GFs and of the CFs, is not fully achieved because they depend on
the external parameters $n$, $T$, $U$, $V$, as well as on the
internal parameters $\mu $, $\kappa ^{(k)}$, $\lambda ^{(k)}$
($k=1,...,2z$). The internal parameters can be self-consistently
computed by using algebra constraints and symmetry requirements
\cite{4}. Within this scheme, in Ref. \cite{4} it has been shown
that the charge correlators $\kappa ^{(k)}$ and $\lambda ^{(k)}$
can be written as a function of only two parameters, $X_{1}$ and
$X_{2}$, in terms of which one may find a solution of the model.
$X_{1}$ and $X_{2}$ are parameters of seminal importance since all
correlators and fundamental properties of the system under study
can be expressed in terms of them. Upon requiring translational
invariance, one finds two equations allowing one to determine
$X_{1}$ and $X_{2}$ as a function of the chemical potential
\cite{4}:
\begin{equation}
\begin{split}
X_{1}& =2e^{\beta \mu }(1-X_{1}-dX_{2})(1+aX_{1}+a^{2}X_{2})^{z-1} \\
& +e^{\beta (2\mu -U)}[2+(d-1)X_{1}-2dX_{2}](1+dX_{1}+d^{2}X_{2})^{z-1}, \\
X_{2}& =e^{\beta (2\mu
-U)}[1+dX_{1}-(2d+1)X_{2}](1+dX_{1}+d^{2}X_{2})^{z-1}
\\
& -2e^{\beta \mu }K^{2}X_{2}(1+aX_{1}+a^{2}X_{2})^{z-1}.
\end{split}
\label{eq36a}
\end{equation}
Here we defined $K=e^{-\beta V}$, $a=(K-1)$ and $d=(K^{2}-1)$. In
turn, the chemical potential $\mu $ can be determined by fixing
the particle density:
\begin{equation}
\begin{array}{c}
n=\frac{\left( X_{1}-2X_{2}\right) \left(
1+aX_{1}+a^{2}X_{2}\right) +2X_{2}\left(
1+dX_{1}+d^{2}X_{2}\right) }{\left( 1-X_{1}+X_{2}\right) +\left(
X_{1}-2X_{2}\right) \left( 1+aX_{1}+a^{2}X_{2}\right) +X_{2}\left(
1+dX_{1}+d^{2}X_{2}\right) }. \label{chem1}
\end{array}
\end{equation}
Equations \eqref{eq36a} and \eqref{chem1} constitute a system of
coupled equations ascertaining the three parameters $\mu$, $X_{1}$
and $X_{2}$ in terms of the external parameters of the model $n$,
$U$, $V$, $T$. Once these quantities are known, all local
properties of the model can be computed.

\section{Results and discussion}
\label{sec_III}

According to the sign of the intersite potential $V$, the
solutions of the self-consistent equations \eqref{eq36a} and
\eqref{chem1} can be remarkably different. Here we study the case
of attractive nearest neighbor interaction $V$ ($V<0$), allowing
the on-site interaction to be both repulsive and attractive. Upon
fixing $V=-1$ and taking $|V|=1$ as the unit of energy, we study
the equations \eqref{eq36a} and \eqref{chem1} for various values
of the external parameters $n$, $T$ and $U$, focusing on the
simpler case $z=3$. By considering higher coordination numbers,
one obtains similar results. In Ref. \cite{4}, upon fixing the
chemical potential at the half filling value $\mu =zV+U/2$, we
have shown that a phase separation exists only for $V<0$. Since
experimentally it is possible to tune the density in a controlled
way, by varying the doping, here we fix the particle density $n$:
the chemical potential will be determined by the system itself
according to the values of the external parameters.

\begin{figure}[t]
\vspace{5mm}
 \centering
 \subfigure[]
   {\includegraphics[scale=0.35]{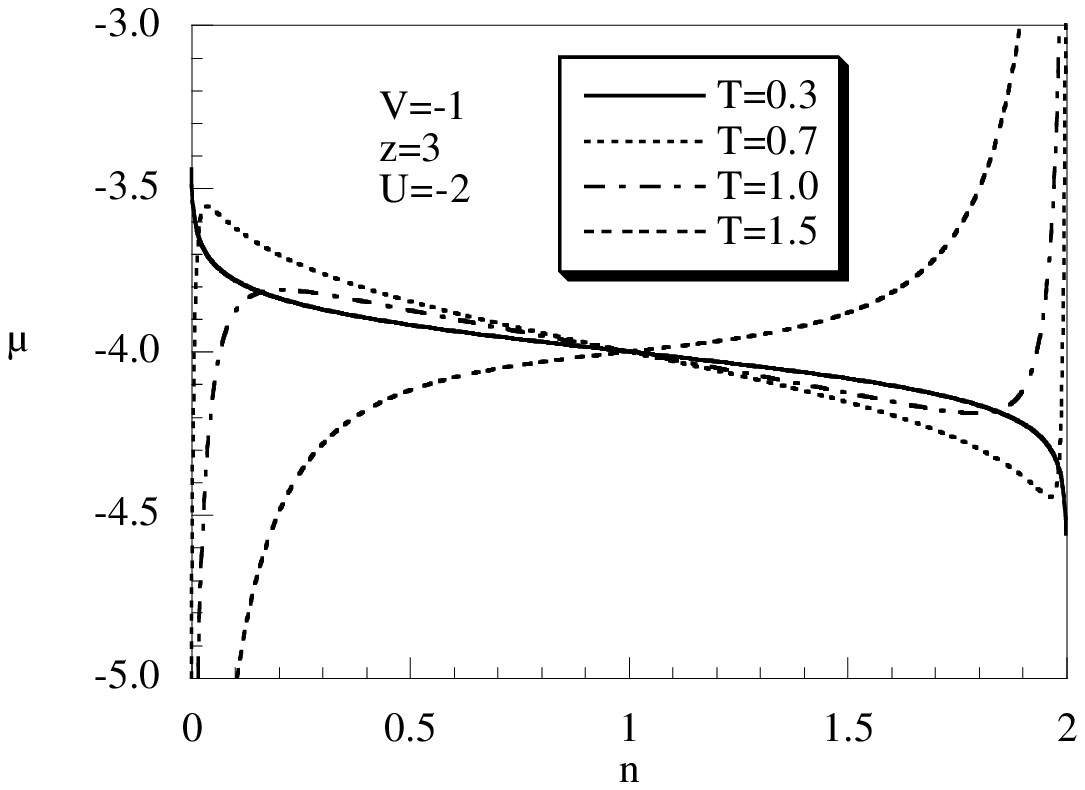}}
 \hspace{1mm}
 \subfigure[]
   {\includegraphics[scale=0.35]{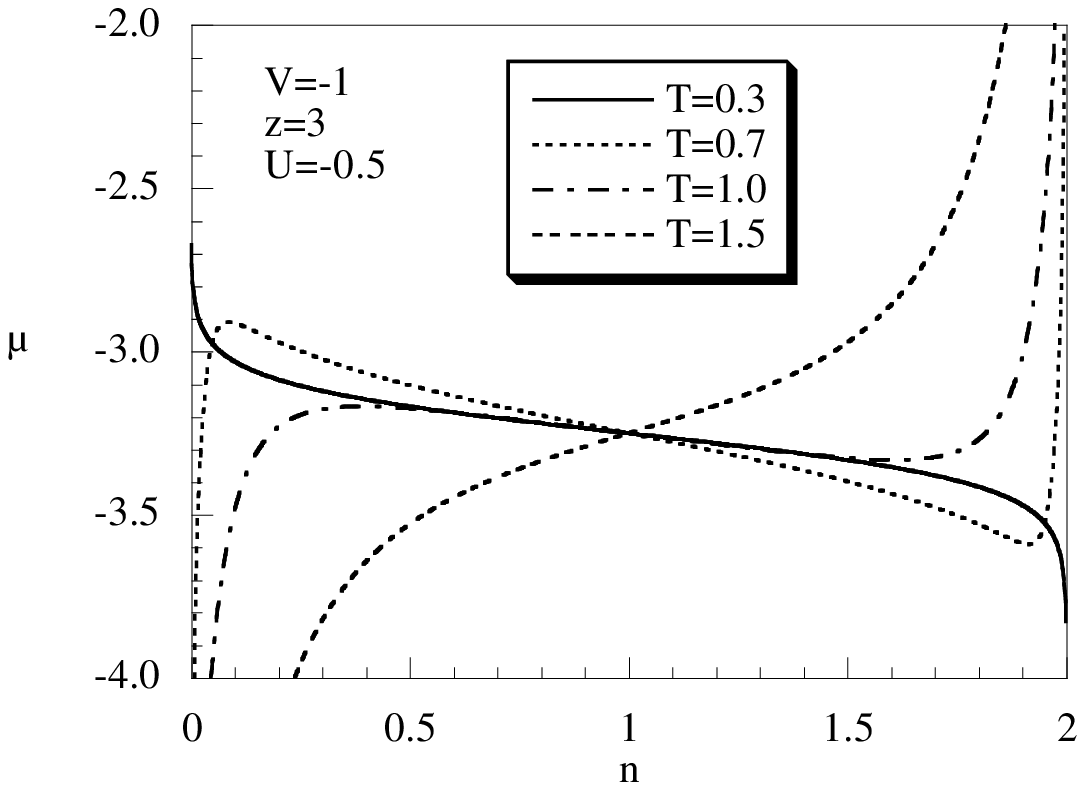}}
\subfigure[]
   {\includegraphics[scale=0.35]{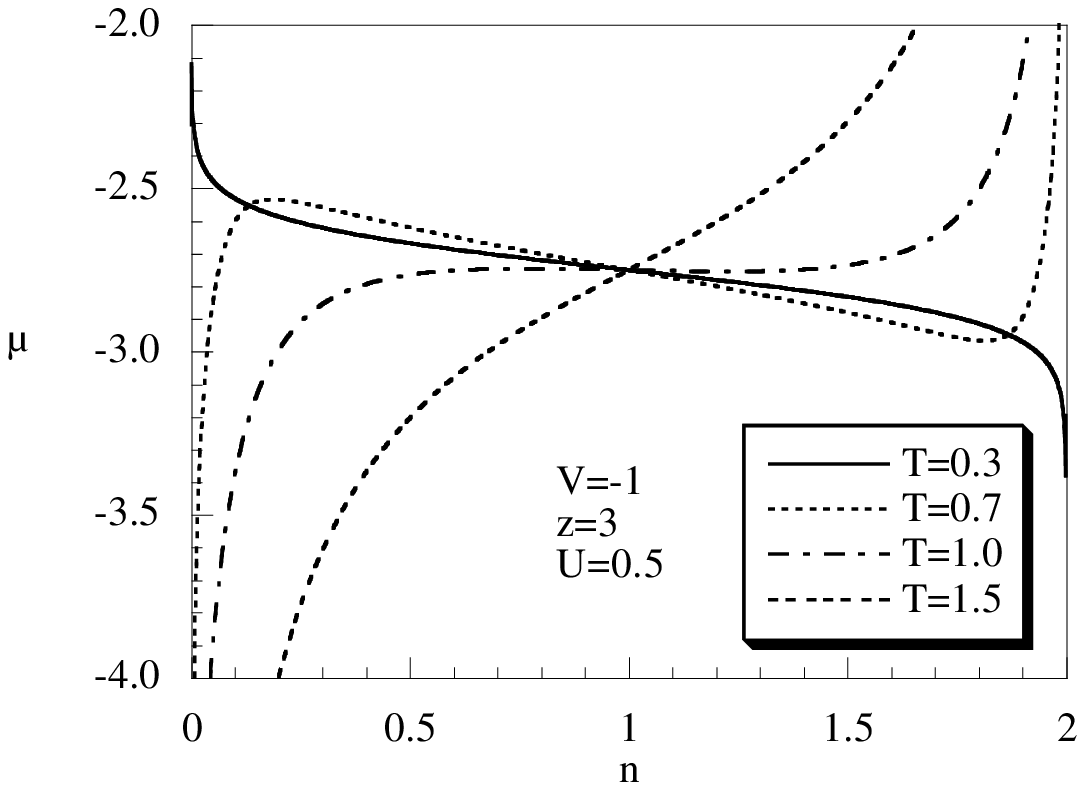}}
   \subfigure[]
   {\includegraphics[scale=0.35]{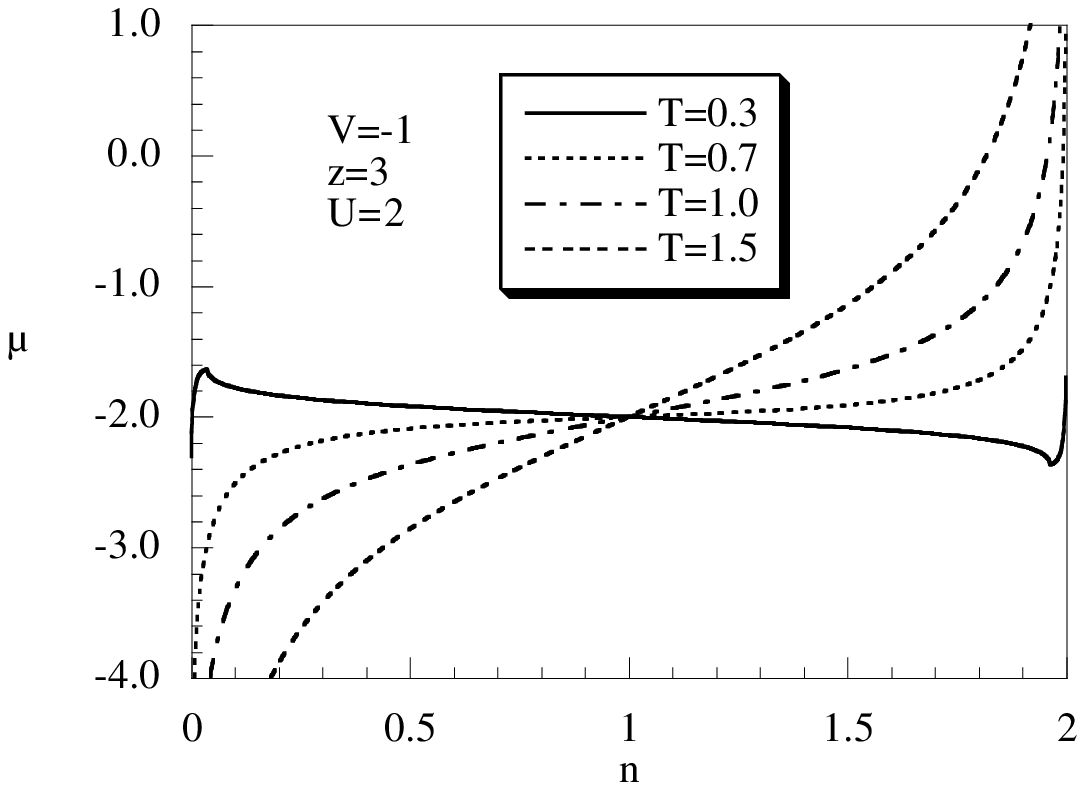}}
\caption{\label{fig2} The chemical potential $\mu$ as a function
of the particle density $n$ for $V=-1$, $z=3$, $T=0.3$, 0.7, 1,
1.5 and for (a) $U=-2$, (b) $ U=-0.5$, (c) $U=0.5$,  and (d)
$U=2$.}
 \end{figure}

In Figs. \ref{fig2}a-d we plot the chemical potential $\mu$ as a
function of the particle density $n$ for both attractive and
repulsive on-site interactions, and for several values of the
temperature. In the high temperature regime, $\mu$ is always an
increasing function of the particle density. By lowering the
temperature, one observes that there is a critical value of the
temperature below which the chemical potential becomes a
decreasing function of $n$ in the interval $ n_{1}\leq n\leq
n_{2}$. The width of the interval depends on both $T$ and $U$.
\begin{figure}[b]
\vspace{5mm}
 \centering
 \subfigure[]
   {\includegraphics[scale=0.35]{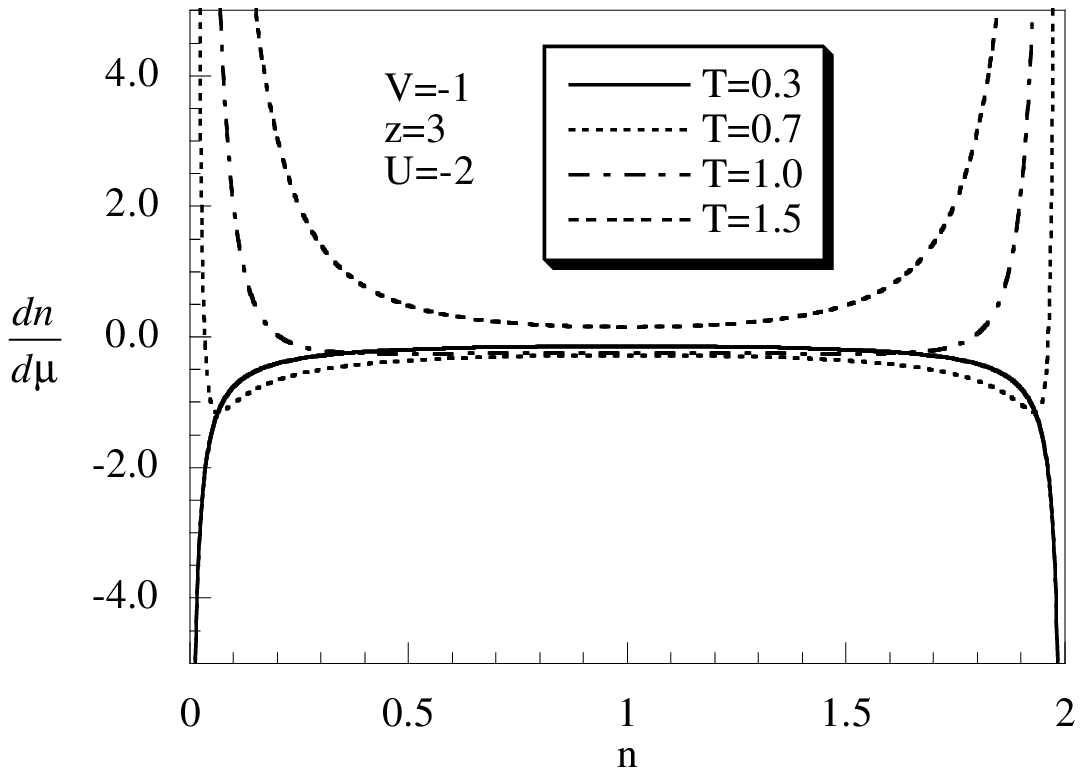}}
 \hspace{1mm}
 \subfigure[]
   {\includegraphics[scale=0.35]{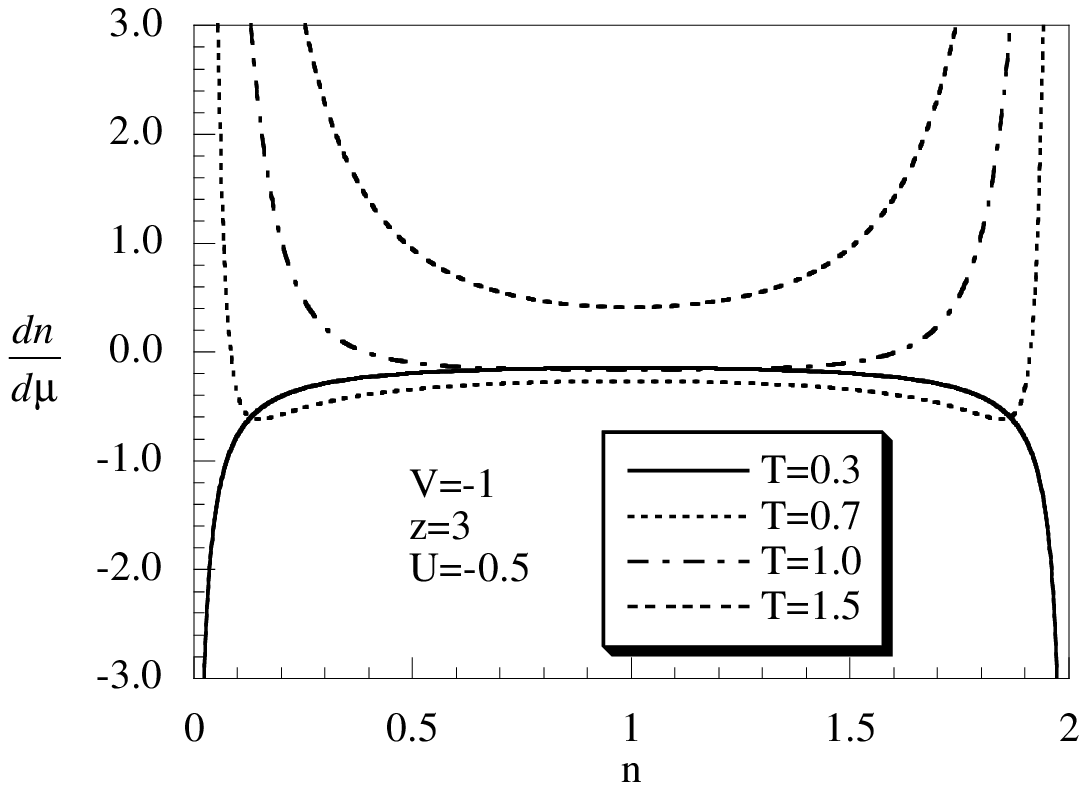}}
    \hspace{30mm}
\subfigure[]
   {\includegraphics[scale=0.35]{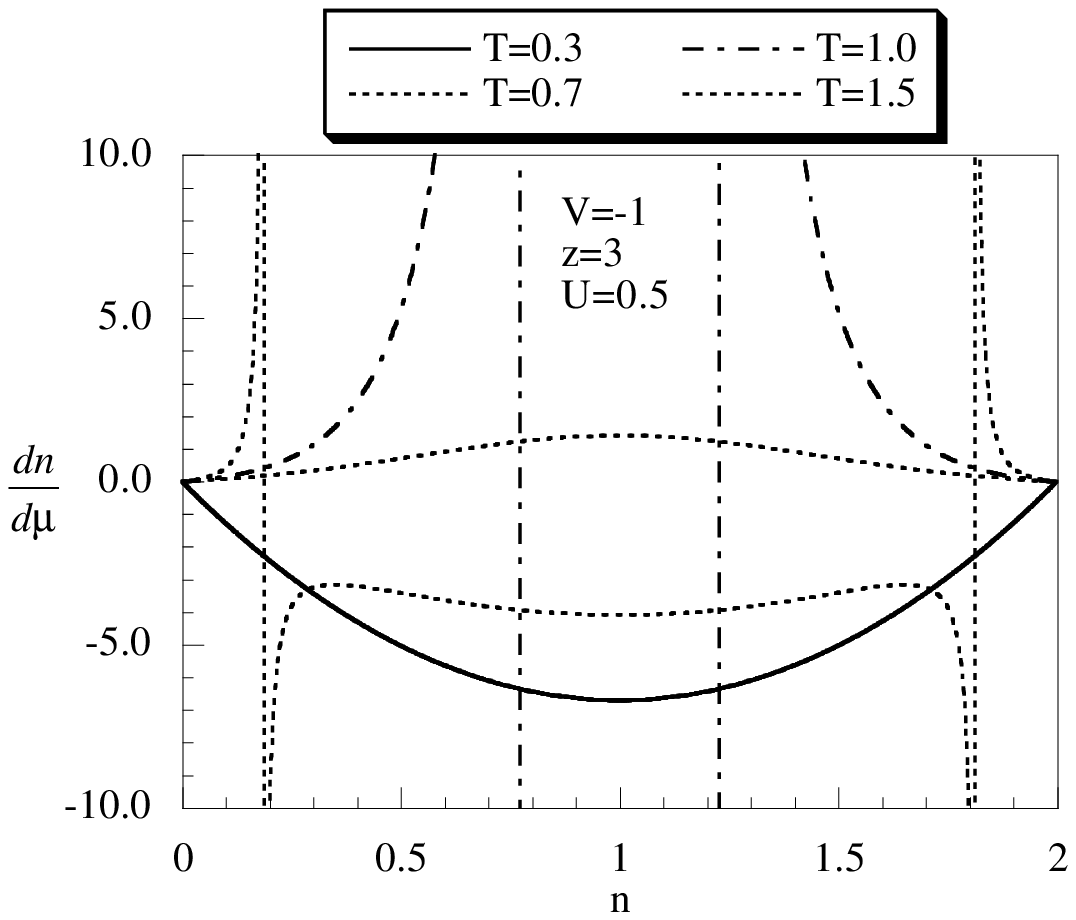}}
   \subfigure[]
   {\includegraphics[scale=0.35]{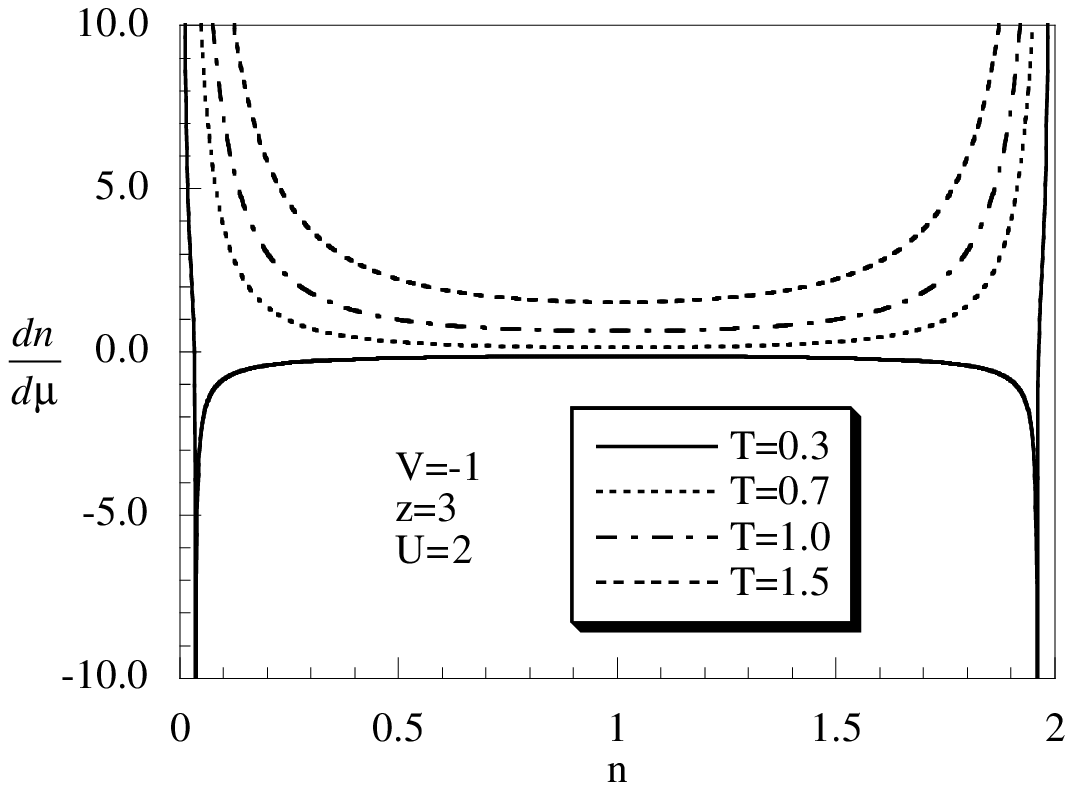}}
\caption{\label{fig3} The derivative $dn/d\protect\mu $ as a
function of the particle density $n$ for $V=-1$, $z=3$, $T=0.3$,
0.7, 1, 1.5 and for (a) $U=-2$, (b) $ U=-0.5$, (c) $U=0.5$, (d)
$U=2$.}
 \end{figure}
In Figs. \ref{fig3} we plot the derivative $dn/d\mu$ as a function
of $n$ for the same values of $T$ and $U$ given in Figs.
\ref{fig2}. One observes that, at high temperatures, $dn/d\mu $ is
a regular function, always positive, for all values of $n$. At low
temperatures, $dn/d\mu $ exhibits a divergence at the critical
points $n_{1}$ and $n_{2}$, is negative in the interval $n_{1}\leq
n\leq n_{2}$, and positive outside. Since the derivative $dn/d\mu
$ is proportional to the thermal compressibility $\kappa = 1/n^{2}
\cdot dn/d\mu$, one immediately infers that $n_{1}\leq n\leq
n_{2}$ is a particle density region where the system is
thermodynamically unstable. A possible solution to this problem is
to abandon the translational invariance assumption made in order
to obtain the self-consistent equations \eqref{eq36a}.
Nevertheless, here we are interested only in describing the
transition to a phase separated regime, and, thus, we shall not
address the problem of finding a thermodynamically stable
solution.
\begin{figure}[t]
\vspace{5mm}
 \centering
 \subfigure[]
   {\includegraphics[scale=0.35]{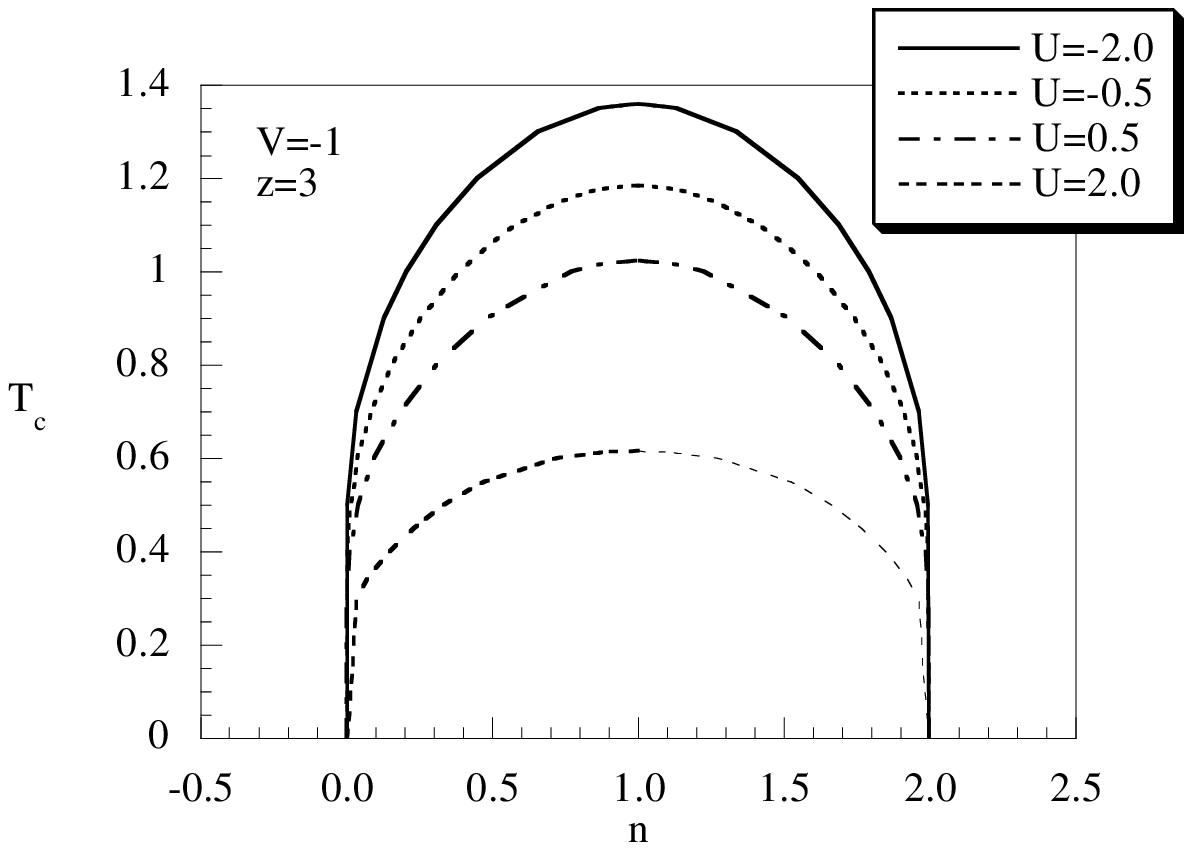}}
    \subfigure[]
   {\includegraphics[scale=0.35]{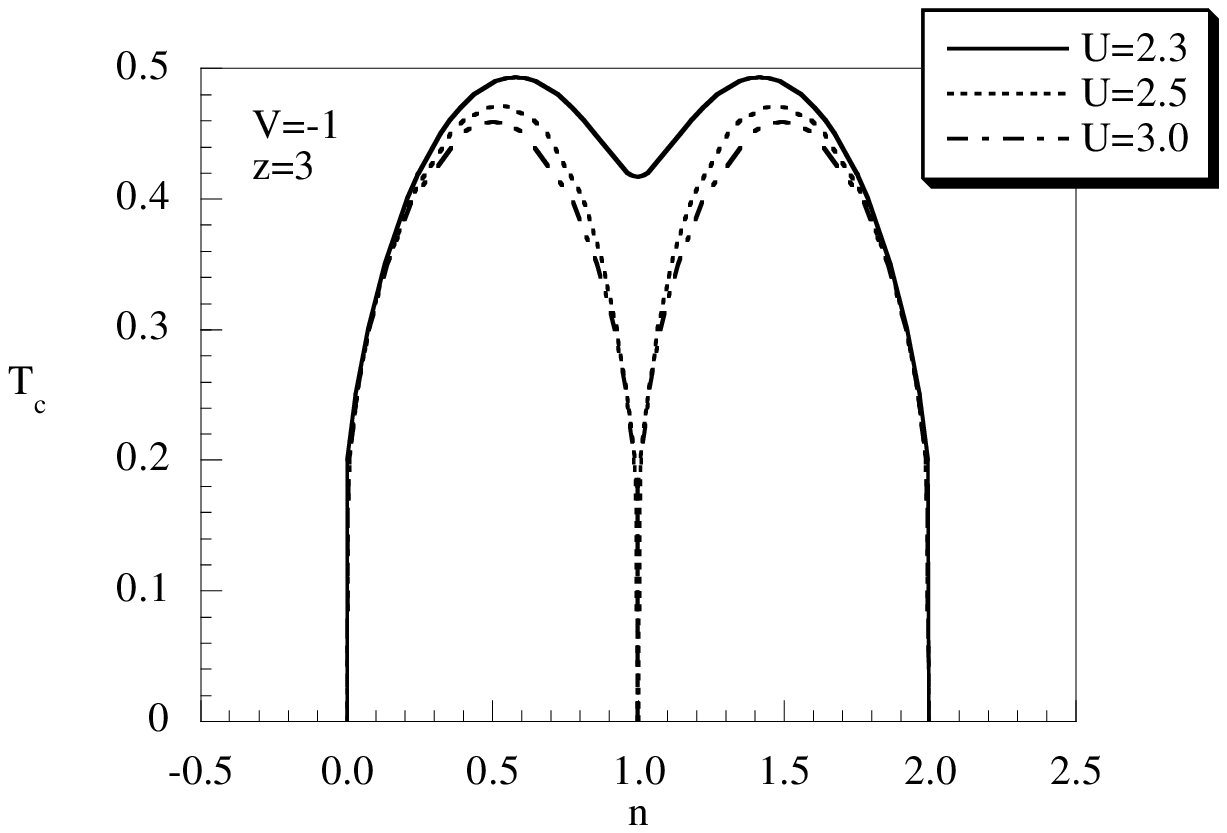}}
 \hspace{1mm}
 \subfigure[]
   {\includegraphics[scale=0.35]{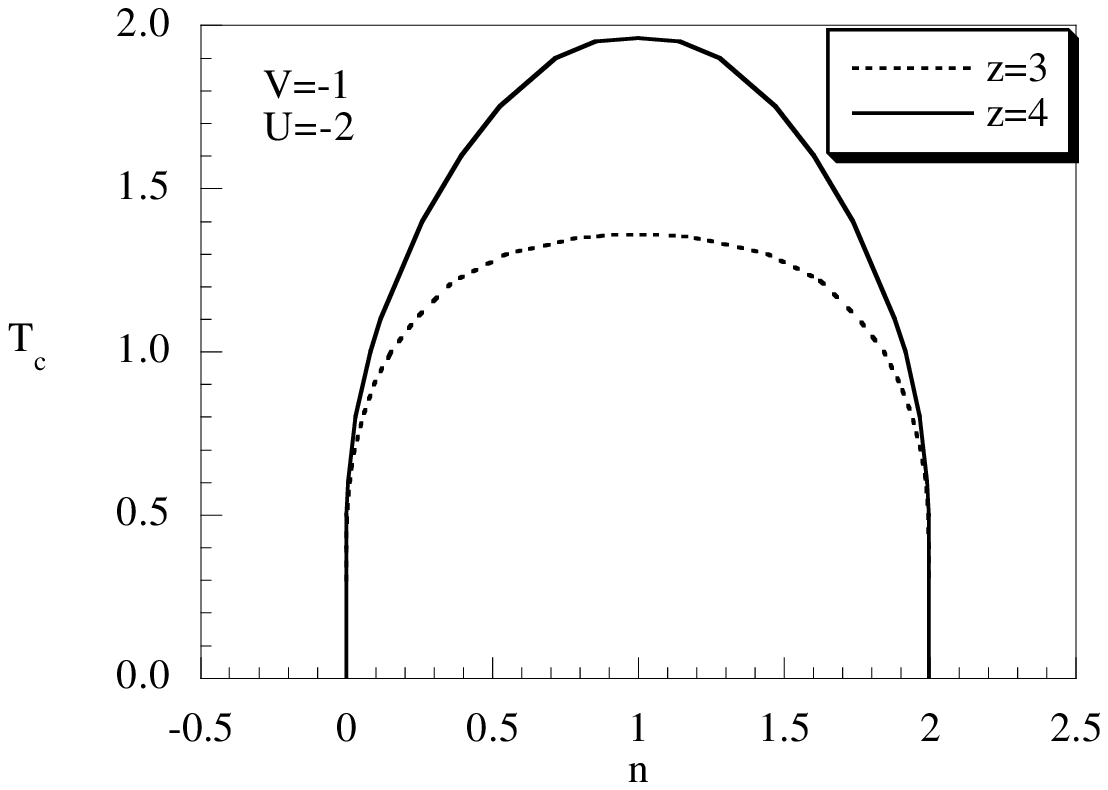}}
\caption{\label{fig4} The critical temperature $T_{c}$ as a
function of the particle density $n$ for $V=-1$, $z=3$ and for (a)
$U=-2$, -0.5, 0.5, 2; (b) $U=2.3$, 2.5, 3. (c) Comparison between
the critical temperatures for $z=3$ and $z=4$.}
 \end{figure}
By studying the temperature dependence of $n_{1}$ and $n_{2}$, one
can obtain the phase diagram in the plane $(n,T)$. For a fixed
value of $n$, there is a critical temperature $T_{c}$ at which the
system undergoes a phase transition from a thermodynamically
stable phase to an unstable one. Below $T_{c}$ the homogeneous
state is not allowed due to a negative compressibility, and a
phase separation occurs. In Figs. \ref{fig4}a-b  we plot $T_{c}$
as a function of the particle density for several values of $U$.
The critical temperature $T_{c}$ and the width of the instability
region $\Delta n=n_{2}-n_{1}$ increase by decreasing $U$: an
attractive on-site potential will favor phase separation and in
particular the clustering of doubly occupied sites. As a function
of the particle density, the critical temperature shows a
lobe-like behavior: it increases by increasing $n$ up to half
filling, where it has a maximum; further augmenting $n$, it
decreases vanishing at $n=2$. A different behavior is observed
when $U>2\left| V\right|$ (see Fig. \ref{fig4}b): the lobe splits
in two parts and $T_c$ increases with $n$ up to quarter filling,
then decreases with a minimum at half filling. The $n=1$ critical
temperature is finite only for $2V<U<U_c$, where $U_c=2.3V$ for
$z=3$. For larger on-site potentials, there is no transition at
half filling. For higher coordination number, one notices an
increase of the critical temperature, as evidenced in Fig.
\ref{fig4}c, where we compare $T_{c}$ at $U=-2$ for $z=3$, 4.
A useful representation of the phase diagram is obtained by
plotting the phase boundary line in the plane $(T_c,U)$. In Fig.
\ref{fig5} we plot the critical temperature $T_{c}$ as a function
of the on-site potential $U$ for several values of $n$. Starting
from large attractive $U$, $T_c$ decreases by increasing $U$ and
increases by increasing $n$, as already evident from Fig.
\ref{fig4}. More interestingly, at half filling one observes a
reentrant behavior with a turning point at $U=U_{c}\approx 2.3V$
(for $z=3$) and at $U=2\left| V\right|$ the critical temperature
vanishes. For particle densities less than half filling, one does
not observe a reentrance in the phase diagram, but rather a
flattening of $T_c$ at a finite value.
\begin{figure}[th]
\vspace{5mm} \centering\includegraphics[scale=0.5]{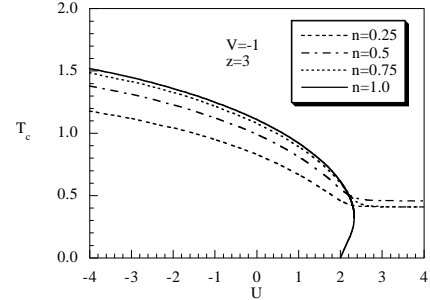}
\caption{\label{fig5} The critical temperature $T_{c}$ as a
function of the on-site potential $U$ for $V=-1$, $z=3$ and for
$n=0.25$, 0.5, 0.75, 1.}
\end{figure}
Although below the critical temperature $T_{c}$ the system is
thermodynamically unstable, the study of the behavior of relevant
parameters below $T_c$, unveils the existence of a critical value
of the on-site potential separating the two possible realizations
of a phase separated region, sketched in Fig \ref{fig1}. In Figs.
\ref{fig6} we plot the double occupancy $D$, the short-range
correlation function $\lambda ^{(1)}$ and the internal energy $E$
as functions of $U$ at half filling and various temperatures.
\begin{figure}[th]
 \centering
 \subfigure[]
   {\includegraphics[scale=0.35]{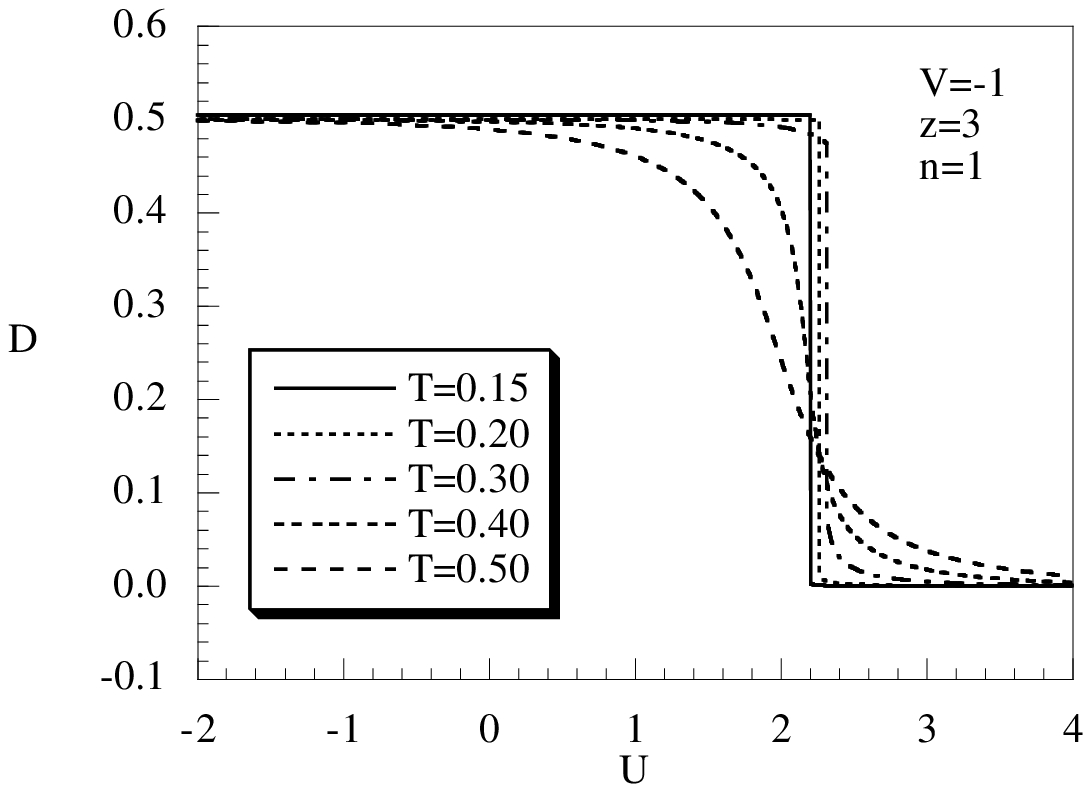}}
 \subfigure[]
   {\includegraphics[scale=0.35]{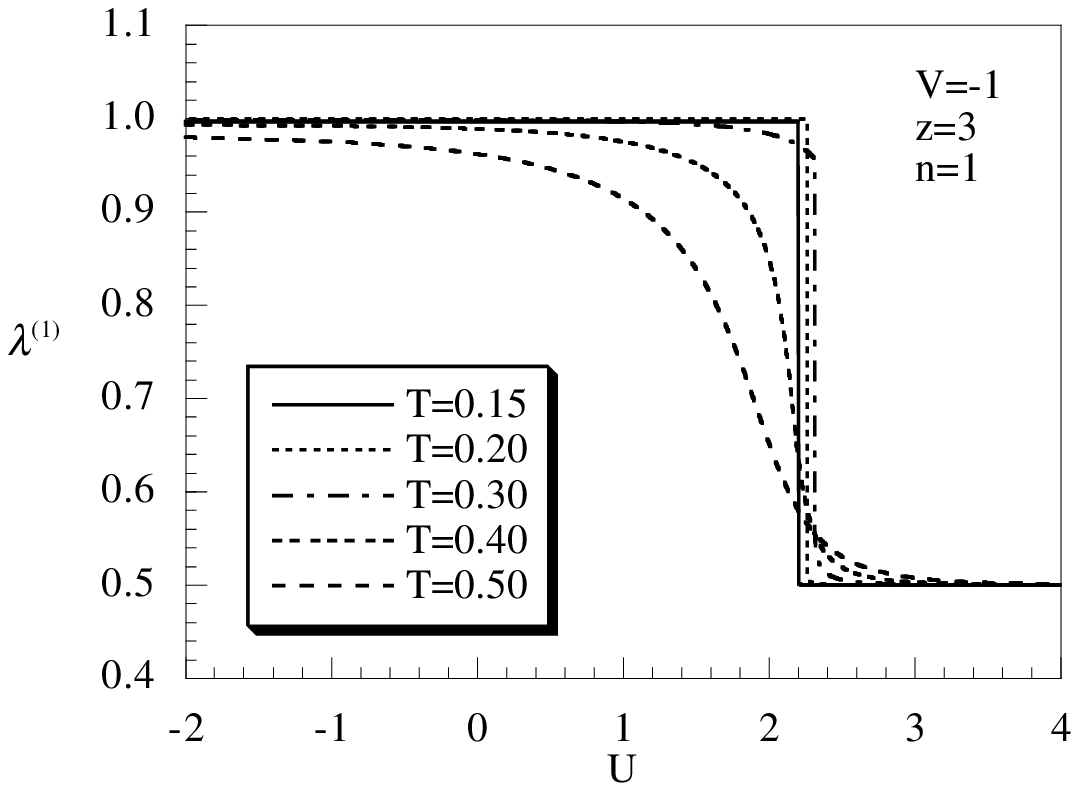}}
   \subfigure[]
   {\includegraphics[scale=0.35]{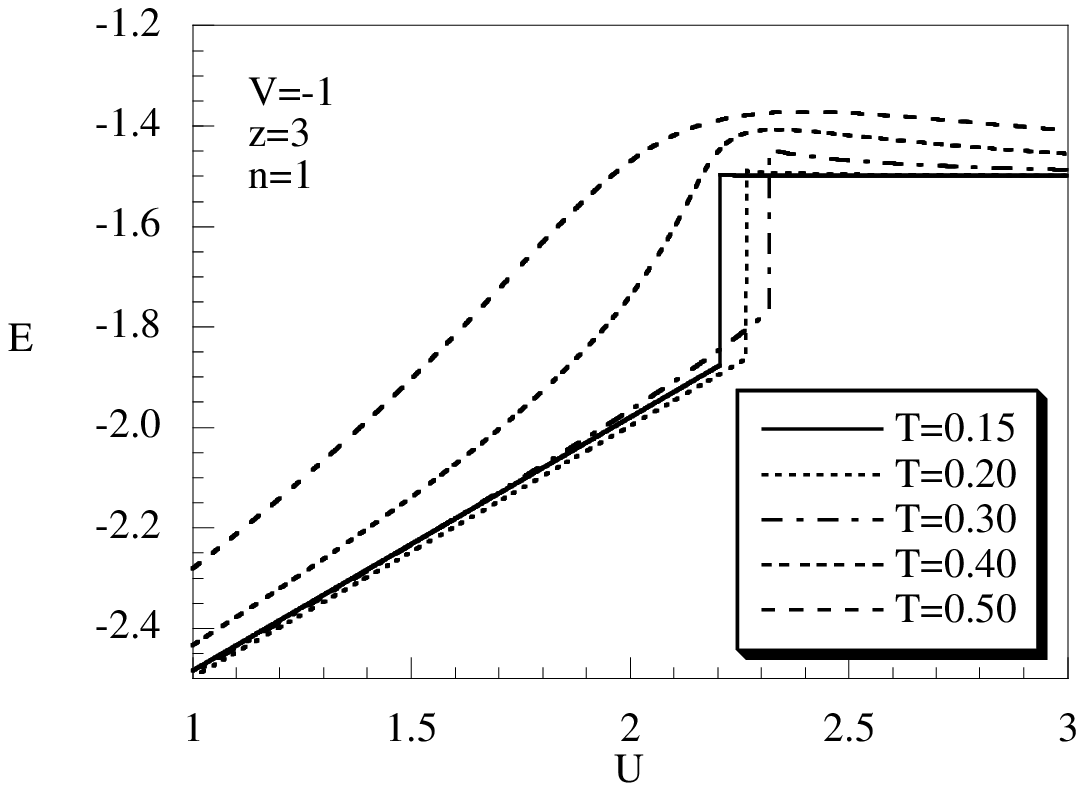}}
\caption{\label{fig6}(a) The double occupancy $D$, (b) the
short-range correlation function $\lambda ^{\left( 1\right)}$ and
(c) the internal energy $E$ as functions of $U$ for $V=-1$, $z=3$,
$ n=1$ and for $T=0.15$, 0.2, 0.3, 0.4, 0.5. }
 \end{figure}
One observes that $D$ and $\lambda^{(1)}$ exhibit two plateaus at
low temperatures. In the limit $T \to 0$ there is a discontinuity
around $U_{c}\approx 2.3V$ (for $z=3$). For $U>U_{c}$, the double
occupancy tends to zero as $T \to 0$, whereas $\lambda ^{(1)}$
tends to $1/2$. The repulsion between the electrons on the same
site and the concomitant nearest neighbor attraction, leads to a
scenario where the electrons tend to cluster together singly
occupying neighboring sites. As a consequence, at half filling all
sites are singly occupied, whereas for $n<1$ one observes two
separated clusters of singly and empty sites. When $U<U_{c}$, one
observes a dramatic increase (step-like) of the double occupancy
and of the short-range correlation function, namely: $D \to 1/2$
and $\lambda ^{(1)}\to 1$. As a consequence, half of the sites are
doubly occupied, and $\lambda ^{(1)}=1$ indicates that the
electrons tend to occupy nearest neighbor sites, arranging to form
large domains occupied, leaving the rest of the lattice empty. For
$U<U_{c}$ one observes also a dramatic decrease of the internal
energy $E$, with a discontinuity as $T\rightarrow 0$ around
$U\approx U_{c}$. In Fig. \ref{fig7} we plot the specific heat $C$
as a function of the temperature $T$ at half filling and for
several values of the on-site potential $U$. One observes that,
for $U\rightarrow U_{c}$, the peak becomes sharper and a
divergence is observed at $U=U_{c}$, confirming the emergence of a
critical point.
\begin{figure}[th]
 \centering
 \subfigure[]
   {\includegraphics[scale=0.35]{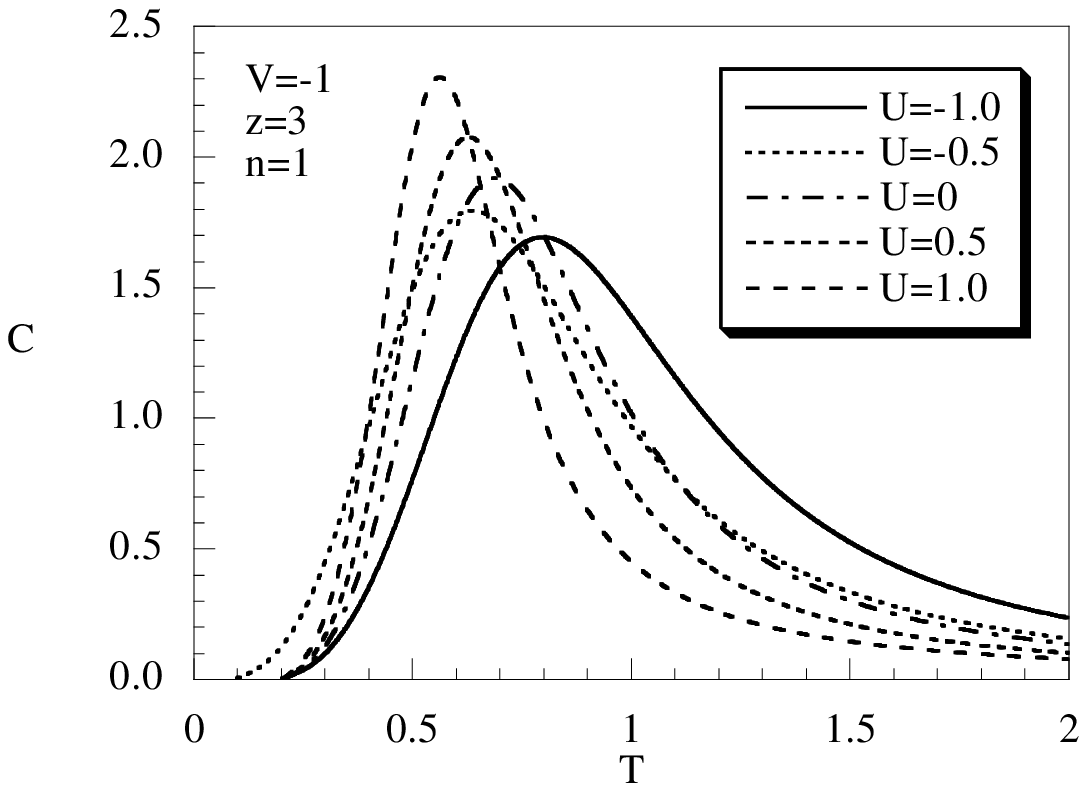}}
 \hspace{1mm}
 \subfigure[]
   {\includegraphics[scale=0.35]{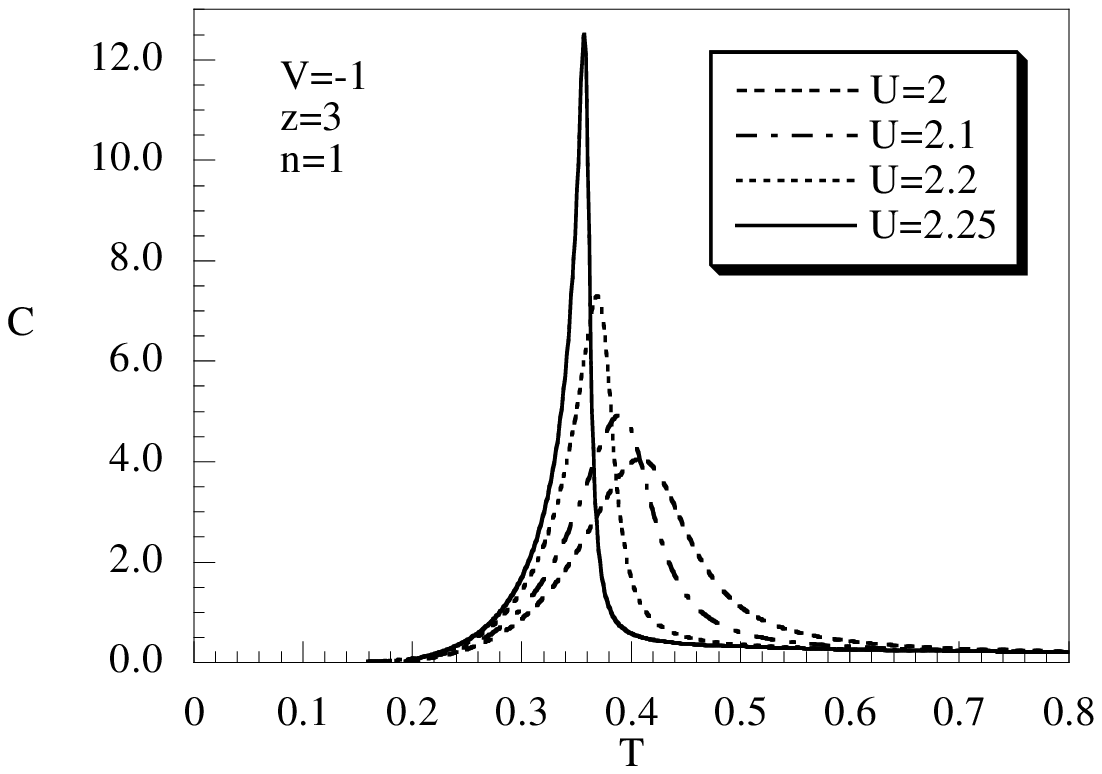}}
\caption{\label{fig7} The specific heat $C$ as a function of the
temperature $T$ for $V=-1 $, $z=3$, $n=1$ and for (a) $U=-1.0$,
-0.5, 0, 0.5, $1.0$; (b) $U=2$, 2.1, 2.2, $2.25$.}
\end{figure}
The same analysis carried out for different values of the particle
density, evidences a similar behavior of $D$, $\lambda ^{(1)}$ and
$E$ as a functions of $U$.

\section{Conclusions}
\label{sec_IV}

In this paper we have obtained the finite temperature phase
diagram of a system of fermions with on-site and nearest neighbor
interactions localized on the sites of the Bethe lattice. The
Hamiltonian describing such a system defines the so-called
extended Hubbard model in the atomic limit. By means of the
equations of motion method, it is possible to exactly solve the
model. For attractive nearest neighbor interaction, by varying the
temperature we found a region of negative compressibility, hinting
at a transition from a thermodynamically stable to an unstable
phase characterized by phase separation. The critical temperature
$T_c$ at which the transition occurs depends on the filling $n$,
on the ratio $U/|V|$ and on $z$. It increases with increasing $z$
and presents a reentrant behavior at half filling for $U/|V|>2$.
Even if the region below $T<T_c$ is thermodynamically unstable,
the study of several thermodynamic quantities allowed us to
determine the critical value of the on-site interaction $U_c$
separating the formation of clusters of singly and doubly occupied
sites. It would be interesting to investigate in detail the
structure of the unstable phase as well as to look for a
thermodynamically stable phase. This could be done by modifying,
for instance, the translational invariance assumption made to
obtain the self-consistent equations \eqref{eq36a} and
\eqref{chem1}.

\end{document}